\documentclass[letterpaper]{emulateapj}
\usepackage{amssymb}
\usepackage{latexsym}
\usepackage{natbib}

\shorttitle{Ly$\alpha$ Heating of Inhomogeneous High-redshift IGM}
\shortauthors{Oklop\v ci\'c, A. \& Hirata, C. M.}

\begin{document}

\title{Lyman-$\alpha$ Heating of Inhomogeneous High-redshift Intergalactic Medium}

\author{Antonija Oklop\v ci\'c\altaffilmark{1} and Christopher M. Hirata\altaffilmark{1,2}}
\email{oklopcic@astro.caltech.edu}
\altaffiltext{1}{Cahill Center for Astronomy and Astrophysics, California Institute of Technology, MC 249-17, 1200 East
California Boulevard, Pasadena, CA 91125 }
\altaffiltext{2}{Center for Cosmology and Astroparticle Physics, Ohio State University, 191 West Woodruff Avenue, Columbus, OH 43210 }

\begin{abstract}
The intergalactic medium (IGM) prior to the epoch of reionization consists mostly 
of neutral hydrogen gas. Lyman-$\alpha$ (Ly$\alpha$) photons produced by early 
stars resonantly scatter off hydrogen atoms, causing energy exchange between the 
radiation field and the gas. This interaction results in moderate heating of the 
gas due to the recoil of the atoms upon scattering, which is of great interest for future
studies of the pre-reionization IGM in the H{\sc\,i} 21 cm line. We investigate the effect of 
this Ly$\alpha$ heating in the IGM with linear density, temperature, and velocity 
perturbations. Perturbations smaller than the diffusion length of photons could 
be damped due to heat conduction by Ly$\alpha$ photons. The scale at which damping 
occurs and the strength of this effect depend on various properties of the gas, 
the flux of Ly$\alpha$ photons and the way in which photon frequencies are 
redistributed upon scattering. To find the relevant length scale and the extent to 
which Ly$\alpha$ heating affects perturbations, we calculate the gas heating rates 
by numerically solving linearized Boltzmann equations in which scattering is treated 
by the Fokker-Planck approximation. We find that (1) perturbations add a small correction 
to the gas heating rate, and (2) the damping of temperature perturbations occurs at scales with 
comoving wavenumber $k \gtrsim 10^4$~Mpc$^{-1}$, which are much smaller than the 
Jeans scale and thus unlikely to substantially affect the observed 21 cm signal.
\end{abstract}


\section {Introduction}
\label{sec:intro}

After recombination of the primordial plasma at redshift $z\approx 1100$ and before 
the epoch of reionization, the baryonic content of the universe was predominantly 
in the form of neutral hydrogen. For this reason, a promising way of probing this 
period in the evolution of the universe is through the observations of the redshifted 
21 cm line of neutral hydrogen, created in the spin-flip transition between the 
two hyperfine levels of the hydrogen ground state (for reviews of the 21 cm physics,
its use in cosmology, and the foreground and calibration challenges see, for example, \citealt{fob06, mw10, prlo12}). There 
are several experiments that are currently in operation, or planned for the near
future, for which a primary objective is observing the redshifted 21 cm signal, such as 
the Low Frequency Array\footnote{www.lofar.org} (LOFAR; \citealt{vanhaarlem13}), the Murchison Widefield 
Array\footnote{www.mwatelescope.org} (MWA; \citealt{lonsdale09}), the Precision Array to Probe EoR\footnote{eor.berkeley.edu} 
(PAPER; \citealt{parsons12}), and the Square Kilometer Array\footnote{www.skatelescope.org}
(SKA; \citealt{rawlings11}). Several pathfinder observations have recently
placed upper limits on the 21 cm perturbation signal at $z=7.7$ \citep{parsons13}, 
$z=8.6$ \citep{paciga13}, and $z=9.5$ \citep{dillon13}, and measurements of the global spectrum have placed
lower limits on the duration of the neutral-to-ionized transition \citep{bowman10}.

The 21 cm signal from high-redshift intergalactic medium (IGM) is sensitive to the 
conditions of the gas, such as its density, ionization fraction, and spin temperature. 
The last can be coupled to the gas kinetic temperature through collisions (in 
environments of sufficiently high density) or via the Wouthuysen-Field effect 
\citep{wout52, field58} in the presence of Lyman-$\alpha$ (Ly$\alpha$) photons. 
Hence, understanding thermal properties of high-redshift IGM is crucial for predicting 
and interpreting the observed 21 cm signal. 

Before the formation of the first sources of radiation, the first stars and galaxies, 
primordial gas was adiabatically cooling with the expansion of the universe -- 
its temperature decreasing with redshift $z$ as $(1+z)^2$. The onset of luminous 
structures dramatically changed that evolutionary track and the universe eventually 
became reheated and reionized. In a complete model of reionization, several different mechanisms 
can affect the temperature and the ionization state of the IGM, such as heating by X-ray 
and UV photons as well as by shocks created in the gravitational collapse of matter. In 
this work, we focus on the microphysics of the IGM heating through its interaction with 
the UV photons. 

Non-ionizing photons emitted by stars can freely travel through mostly neutral 
high-redshift IGM until they are redshifted by the expansion of the universe into 
a resonant frequency of one of the atomic species present in the IGM, at which 
point they resonantly scatter with atoms. This scattering can occur far away 
from sources of radiation. Since hydrogen is the most abundant element in the 
universe and is mostly in its ground state at high redshifts prior to reionization, 
the most significant resonance is the Ly$\alpha$ transition between the ground 
state and the first excited state of hydrogen ($\lambda_{\alpha} = 1216 \ \AA$, 
$\nu_{\alpha} = 2.47\times 10^{15}$~Hz). There are two types of Ly$\alpha$ photons 
that need to be taken into account: Photons emitted with frequencies between Ly$\alpha$
and Ly$\beta$ can redshift directly into the Ly$\alpha$ resonance, whereas photons 
of higher energies (between Ly$\gamma$ and the Lyman limit) can fall into one of 
the higher Lyman resonances, from which they can cascade into Ly$\alpha$ (see, for example,
\citealt{prfu06, hirata06}). To distinguish these two types of photons, we 
call the first kind the \textit{continuum photons} and the second kind the 
\textit{injected photons}.

The resonant scattering of Ly$\alpha$ photons with hydrogen atoms causes transfer of 
energy from the radiation field to the gas due to atomic recoil, causing 
a change in the kinetic temperature of the gas (\citealt{mmr97}; see also 
\citealt{cme04,fupr06,meiksin06,cisal07}). This energy exchange leads to 
a drift of photons from higher to lower frequencies. Another contribution to 
frequency drift, one that is present regardless of scattering, comes from the 
Hubble expansion of the universe. Scattering also causes a diffusion of photons 
in frequency space due to a Maxwellian distribution of atomic velocities in the 
gas. Photons on the red side of the Ly$\alpha$ line center mostly scatter with 
atoms moving toward them, and because of the Doppler shift, the frequency of these 
photons is higher in the frame of the atom; in other words, it is closer to the 
line center and the resonant frequency. The opposite occurs for the photons on 
the blue side of the line -- they preferentially scatter off atoms moving away 
from them. Frequent scattering between atoms and photons brings them closer to 
statistical equilibrium, reducing the average energy exchange per scattering 
\citep{cme04}. Fluctuations in the temperature and density of the gas, as well 
as gradients in its velocity, can change the Ly$\alpha$ scattering rates 
(\citealt{higg09}, \citeyear{higg12}) and thereby affect the heating of the 
gas by Ly$\alpha$ photons. Therefore, it is of great interest to understand how 
the theory of IGM heating by Ly$\alpha$ photons extends to the case of a realistic, 
inhomogeneous universe -- whether the heating rate is just slightly modified by 
the perturbations, or whether effects such as thermal conduction can become important.

In this study we investigate the Ly$\alpha$ heating of hydrogen gas with underlying 
perturbations in the density, temperature, and baryonic velocity. We assume that these 
perturbations are small and consider their contribution only to linear order. We find 
that, as a consequence of perturbations, the gas heating rate can be altered by a few 
percent compared to the heating rate in a homogeneous medium. Of particular interest 
are perturbations with scales comparable to or smaller than the diffusion length of 
Ly$\alpha$ photons. For perturbations on these scales, photons can interact with 
hydrogen atoms located in regions with different properties than where the photons 
originated from, changing the gas heating rate. This process can therefore be viewed 
as thermal conduction between regions of different temperatures, which could lead to 
the damping of perturbations. The spatial redistribution of photons depends in a complicated 
way not only on the properties of the gas but also on the rate of frequency redistribution 
of photons since the scattering cross section (and hence spatial diffusion coefficient) 
varies by many orders of magnitude over the frequency range of interest. Hence, as it is 
difficult to make a simple estimate of the diffusion length of photons, we solve the 
problem numerically. Our results show that the scale at which perturbations start to 
counteract the mean effect, and hence damp the perturbations, corresponds to a wavenumber 
of $k\sim 10^4$~Mpc$^{-1}$ (comoving). That length scale is roughly two orders of magnitude 
smaller that the Jeans scale.

This paper is structured as follows: At the beginning of Section \ref{sec:form}, 
we introduce the notation and outline the formalism that is used in our analysis. We 
continue by describing the radiative transfer equations and the resulting radiation 
spectra. Heating rate calculations for the continuum and injected photons are described 
in Section \ref{sec:hrate}. We present our results in Section \ref{sec:results} and 
finally discuss and summarize our conclusions in Section \ref{sec:conclusions}.
 
Throughout this paper we assume the following values of the relevant cosmological parameters, 
obtained by the \citet{PlanckPar}: $H_0 = 67.3$~km~s$^{-1}$~Mpc$^{-1}$, $\Omega_{\Lambda} = 0.685$, 
$\Omega_m = 0.315$.

\section{Formalism}
\label{sec:form}

Our formalism is based on following the time evolution of photon phase-space distribution, 
which is governed by the Boltzmann equation. The approach is similar to that developed for 
studying the cosmic microwave background \citep{ma95}, except that in our steady-state case, 
the nontrivial variable is frequency rather than the time. In our calculation, we neglect 
polarization since its effect on the radiation intensity is expected to be small, and to 
include it in the calculation would require tracking twice as many variables.
 
We start the analysis by considering the phase-space density of photons of frequency $\nu$, 
located at coordinate $\textbf{x}$ and propagating in direction $\hat{n}$, given by the 
occupation number $f_{\nu}(\textbf{x},\hat{n})$. To simplify our equations, from now on we 
omit writing $(\textbf{x},\hat{n})$ explicitly, although we assume such dependence in 
calculations. The occupation number $f_{\nu}$ consists of two parts, the mean isotropic part 
$\overline{f}_{\nu}$, and direction-dependent perturbations $\delta f_{\nu}$: 
\begin{eqnarray}
f_{\nu} = \overline{f}_{\nu} + \delta f_{\nu} \ \mbox{.}
\end{eqnarray}
The scale of perturbation is determined by its wavenumber $k$. In the equations given throughout 
this paper, $k$ is used to denote the physical wavenumber, rather than comoving, which simplifies 
expressions. We convert to the comoving wavenumber only at the end when we report the final results 
and present them in figures. We assume that all perturbations are small and linear. So we can 
treat them independently, since in linear perturbation theory, different $k$-modes are decoupled 
form each other. The contribution of a single $k$-mode to $\delta f_{\nu}$ can be expanded in a 
series of Legendre polynomials with 
coefficients $\delta f_{l\nu}$:
\begin{eqnarray}
\delta f_{\nu} =  \sum_{l=0}^{\infty} i^l (2l+1)\delta f_{l\nu} P_l(n_3)e^{ikx_3} \mbox{,}
\end{eqnarray}
where $n_3$ is the projection of unit vector $\hat{n}$ onto $x_3$ axis.

The IGM can be described by its mean number density $\overline{n}$ and the mean gas kinetic 
temperature $\overline{T}$. However, for an inhomogeneous medium, the density and temperature
fields are given by:
\begin{equation}
n(\textbf{x}) = \overline{n}(1+\delta_n e^{ikx_3})
\end{equation}
and
\begin{equation}
T(\textbf{x}) = \overline{T}(1+\delta_T e^{ikx_3}) \ \mbox{,}
\end{equation}
where $\delta_n$ and $\delta_T$ are dimensionless parameters describing the amplitudes of density
and temperature perturbations, respectively.

We treat the photon field in the rest frame of the baryons - not the comoving frame - because a 
photon will resonantly scatter with an atom if the photon frequency matches the resonant 
frequency in the atom's rest frame. In that frame, the overall mean velocity of atoms vanishes.
We introduce linear perturbations in the baryonic velocity:
\begin{equation}
\textbf{v}(\textbf{x}) = \delta_v e^{ikx_3}\textbf{e$_3$} \ \mbox{,}
\end{equation}
where $\delta_v$ is taken to be imaginary. Velocity divergence is then given by
\begin{equation}
\Theta (\textbf{x}) = \nabla \textbf{v}(\textbf{x}) = ik\delta_v e^{ikx_3} =  \delta_{\Theta} e^{ikx_3}  \mbox{.}
\end{equation}

\subsection{Radiative Transfer}
\label{sec:radtrans}

Time evolution of the photon distribution function is governed by the Boltzmann equation
\begin{equation}
\frac{\partial f_{\nu}}{\partial t} + \frac{d \nu}{d t}\frac{\partial
 f_{\nu}}{\partial \nu} + \frac{dx_i}{dt}\frac{\partial f_{\nu}}{\partial x_i} + \frac{dn_i}{dt}\frac{\partial f_{\nu}}{\partial n_i}  = \frac{\partial f_{\nu}}{\partial t}\bigg|_{coll} \ \mbox{.}
\label{eq:Boltzmanneq}
\end{equation}
The left-hand side of the equation describes the free steaming of photons, and the collision 
term is on the right-hand side. To keep our analysis linear in small quantities, we ignore the 
last term on the left, which represents gravitational lensing, because both factors are of the 
first order in perturbations, making the entire term second order. Therefore, our linearized 
collisionless equation is
\begin{equation}
\frac{df_{\nu}}{dt} \approx \frac{\partial f_{\nu}}{\partial t} + \frac{d \nu}{d t}\frac{\partial
 f_{\nu}}{\partial \nu} + \frac{dx_i}{dt}\frac{\partial f_{\nu}}{\partial x_i} \ \mbox{.}
\label{eq:Bolcollisionless}
\end{equation}
Next, we evaluate different terms of this equation in the Newtonian gauge.
The second term on the right side includes the time change in the photon frequency due to the 
expansion of the universe and the relative motion of the baryons because the frequency in Equations 
(\ref{eq:Boltzmanneq}) and (\ref{eq:Bolcollisionless}) is defined relative to the baryons, rather than 
to an observer fixed in Newtonian coordinates. We ignore contributions of the time derivative 
of metric perturbation because the metric potential is negligible compared to the subhorizon 
perturbations in the baryons that we are considering in this analysis.\footnote{Using the basic equations 
of the linear perturbation theory, it can be shown that the amplitude of baryonic perturbations
is proportional to $\left(kc/H\right)^2 \Phi $, where $\Phi$ is the metric potential. For subhorizon 
perturbation, the wavenumber $k$ is much larger than $H/c$, making $\Phi$ negligible 
compared to perturbations in the baryons. Similarly, it can be shown that the baryonic velocity
is proportional to $\left(kc/H\right) \Phi $. Hence, we can neglect the gravitational redshift/blueshift 
because it is small compared to the redshift/blueshift due to peculiar motions of the baryons.}
The third term is proportional to the gradient of $f_{\nu}$ and the only contribution to that term comes from 
the perturbative part of the photon distribution function $\delta f_{\nu}$. The accompanying factor is 
just the velocity of photons in the direction of the $x_3$ axis, which is equal to $cn_3$.

Focusing for now only on the collisionless Boltzmann equation, we set it equal to zero and get the following
expression for the time evolution of the photon occupation number in the free-streaming (collisionless) case
\begin{eqnarray}
\frac{\partial f_{\nu}}{\partial t}\bigg|_{\rm fs} &=& \left(H\nu + n_3^2\nu e^{ikx_3}\delta_{\Theta} \right)\frac{\partial f_{\nu}}{\partial \nu}
\nonumber\\ && - ikcn_3\sum_{l=0}^{\infty}i^l(2l+1) \delta f_{l\nu} P_l\left(n_3\right) e^{ikx_3} \ \mbox{,}
\nonumber\\ &&
\end{eqnarray}
where $H$ is the Hubble parameter.

Using the basic properties and recurrence relations of Legendre polynomials, we find the 
expressions for each multipole order
\begin{equation}
f_{l\nu}=\frac{1}{2i^l}\int f_{\nu} P_l(n_3) dn_3
\label{eq:legpol}
\end{equation}
and its time derivative
\begin{eqnarray}
\nonumber \dot{f}_{l\nu} \bigg|_{\rm fs} &=& H\nu\left(\frac{\partial \overline{f}_{l\nu}}{\partial \nu} + e^{ikx_3}\frac{\partial \delta f_{l\nu}}{\partial\nu} \right) \\
\nonumber &&+ \nu\delta_{\Theta}e^{ikx_3}\frac{l(l-1)}{(2l-1)(2l+1)} \frac{\partial \overline{f}_{(l-2)\nu}}{\partial \nu}\\
\nonumber &&+ \nu\delta_{\Theta}e^{ikx_3}\frac{(l+1)^2(2l-1)+l^2(2l+3)}{(2l-1)(2l+1)(2l+3)}\frac{\partial \overline{f}_{l\nu}}{\partial \nu}  \\
\nonumber &&+ \nu\delta_{\Theta}e^{ikx_3}\frac{(l+2)(l+1)}{(2l+1)(2l+3)}\frac{\partial \overline{f}_{(l+2)\nu}}{\partial\nu}\\
 &&- \frac{kc e^{ikx_3}}{2l+1} \left[l\delta f_{(l-1)\nu}-(l+1)\delta f_{(l+1)\nu}\right] \  \ \mbox{.}
\label{eq:fsfull}
\end{eqnarray}
Since we assume that $\overline{f}_{\nu}$ is isotropic, only the monopole term
($\overline{f}_{0\nu}$) is nonzero, and the above expression is therefore greatly simplified for most multipole 
orders. More specifically, the second line is nonzero only for $l=2$, and the third line contributes only 
to the equation for $l=0$, whereas the fourth line vanishes for all values of $l$.

The right side of the full Boltzmann equation (Equation (\ref{eq:Boltzmanneq})) describes the 
change in the photon occupation number due to collisions with atoms. It consists of two terms, 
one describing photons scattered into the phase-space element of interest and the other
describing the outgoing photons \citep{rybdell94}:
\begin{eqnarray}
\nonumber \frac{\partial f_{\nu}}{\partial t}\bigg|_{\rm coll} &=&  \int n_H \sigma\left(\nu^{\prime}\right)cf_{\nu^{\prime}}(\hat{n}^{\prime}) R(\nu^{\prime}\hat{n}^{\prime}, \nu \hat{n}) d\nu^{\prime} d^2\hat{n}^{\prime}\\
&-& n_H \sigma\left(\nu\right)c f_{\nu}\left(\hat{n}\right) \mbox{,}
\label{eq:coll}
\end{eqnarray}
where $n_H$ is the number density of hydrogen atoms and $\sigma(\nu)= \sigma_0\Phi(\nu)$ is the 
collisional cross section at frequency $\nu$ given by the cross section at the line center 
$\sigma_0$ and the Voigt profile $\Phi(\nu)$:
\begin{eqnarray}
\sigma(\nu) = \frac{\pi e^2}{m_e c}\frac{f_{12}}{\Delta\nu_D}\frac{a}{\pi^{3/2}}\int_{-\infty}^{+\infty} dy\frac{e^{-y^2}}{(x-y)^2+a^2} \ \mbox{,}
\end{eqnarray}
where $a=A_{21}/(8\pi\Delta \nu_D)$ is the Voigt parameter and $A_{21}=6.25\times 10^8$~s$^{-1}$ 
is the Einstein coefficient of spontaneous emission for the Ly$\alpha$ transition. The Doppler width 
of the line is given by
\begin{equation} 
\Delta \nu_D = \nu_{\alpha}\sqrt{\frac{2k_BT}{m_Hc^2}}
\end{equation} 
 and $x$ is used to denote the offset from the line center
\begin{equation}
x=\frac{\nu-\nu_{\alpha}}{\Delta\nu_D} = \frac{\Delta \nu}{\Delta \nu_D} \ \mbox{.}
\end{equation}
The probability that a photon of frequency $\nu'$, propagating in the direction of $\hat{n}^{\prime}$, 
will be redistributed upon scattering into a photon of frequency $\nu$, propagating in the direction 
of $\hat{n}$, is represented by $R(\nu^{\prime}\hat{n}^{\prime}, \nu \hat{n})$. It can be decomposed 
into a series of Legendre polynomials in terms of the scattering angle, the cosine of which is given 
by the dot product of 
$\hat{n}$ and $\hat{n}^{\prime}$:
\begin{equation}
R(\nu^{\prime}\hat{n}^{\prime}, \nu \hat{n}) = \frac{1}{4\pi}\sum_{l}R(l;\nu, \nu^{\prime})P_l(\hat{n}\cdot\hat{n}^{\prime}) \ \mbox{.}
\end{equation} 
Plugging this into Equation (\ref{eq:coll}) and using the obtained expression in the time derivative 
of Equation (\ref{eq:legpol}) gives
\begin{eqnarray}
\nonumber \frac{\partial f_{l\nu}}{\partial t}\bigg|_{\rm coll} &=&  \frac{n_H c}{2i^l}\int \sigma\left(\nu^{\prime}\right)f_{\nu^{\prime}} d\nu^{\prime} d\phi d(\cos{\theta}^{\prime})d(\cos{\theta})\times\\
\nonumber &&  P_l(\cos{\theta})\frac{1}{4\pi}\sum_{l}R(l; \nu^{\prime}, \nu)P_l(\hat{n}\cdot\hat{n}^{\prime}) \\
&&- n_H \sigma\left(\nu\right)c f_{l\nu}  \mbox{.}
\end{eqnarray}
Using the spherical harmonic addition theorem 
\begin{equation}
P_l(\hat{n}\cdot\hat{n}^{\prime}) = \frac{4\pi}{2l+1}\sum_{m=-l}^l Y_{lm}^*(\theta^{\prime}, \phi^{\prime})Y_{lm}(\theta, \phi) \ \mbox{,}
\end{equation}
the relation between spherical harmonics and Legendre polynomials
\begin{equation}
Y_{lm}(\theta, \phi) = \sqrt{\frac{(2l+1)(l-m)!}{4\pi(l+m)!}}P_l^m(\cos{\theta})e^{im\phi} \ \mbox{,}
\end{equation}
and the orthogonality of Legendre polynomials, we get
\begin{eqnarray}
\nonumber \frac{\partial f_{l\nu}}{\partial t}\bigg|_{\rm coll} &=& \int n_H \sigma\left(\nu^{\prime}\right)cf_{l\nu^{\prime}} R(\nu^{\prime}, \nu)\delta_{l0} d\nu^{\prime}  \\
&&- n_H \sigma\left(\nu\right)c f_{l\nu}  \mbox{.}
\end{eqnarray}
Here we assume that the emission of photons is isotropic (i.e., nonzero only for $l=0$), which is a 
reasonable assumption for a medium that is optically thick at the resonant frequency. The outgoing 
part of the collision term is direction dependent in the case of an inhomogeneous medium. Hence, we 
keep that term for all multipole orders $l$. For multipoles with $l>0$, this term dominates and causes 
their attenuation. For $l=0$, on the other hand, the incoming and outgoing terms nearly cancel, which 
is why the monopole equation needs to be treated differently.

The redistribution function $R(\nu^{\prime}, \nu)$ is generally very complicated. However, it can 
be simplified if the radiation spectrum changes smoothly on the scale of a typical change of the 
photon frequency in a single scattering, which is on the order of $\Delta \nu_D$. If this condition 
is satisfied, we can use the Fokker-Planck approximation in which scattering is treated as diffusion 
in frequency space. Using the result of \cite{ryb06}, the collision term for the monopole order 
then becomes
\begin{eqnarray}
\frac{\partial f_{0\nu}^{coll}}{\partial t} &=& \frac{1}{\nu^2}\frac{\partial}{\partial \nu}\left[ \nu^2D_{\nu}\left(\frac{\partial f_{0\nu}}{\partial \nu} + \frac{hf_{0\nu}}{k_BT}\right)\right] + \Psi \mbox{,}
\label{eq:fpap}
\end{eqnarray}
where $\Psi$ is the photon source term describing photons injected with a frequency distribution
that can be approximated by a delta function around the Ly$\alpha$ frequency. This term is used 
for describing the injected photons, whereas it vanishes in the case of the continuum photons. 
The parameter $D_{\nu}$ (in units of Hz$^2$ s$^{-1}$) is the frequency diffusivity, given by
\cite{hirata06}:
\begin{equation}
D_{\nu} = \frac{3k_BT}{m_H}\gamma n_H x_{HI}c\Phi(\nu)\ \mbox{.}
\end{equation}
Here $\gamma = 50$~MHz is the half width at half maximum of the Ly$\alpha$ resonance, $x_{HI}$ is
the neutral fraction of hydrogen, and $m_H$ is the mass of the hydrogen atom.

Equating the result for the free streaming and the collision term, and assuming that a steady state 
($\partial f_{\nu}/\partial t =0$) has been reached, gives the full expression for the radiative
transport. We can write an equation for each multipole order separately, producing an infinite series 
of coupled differential equations -- the Boltzmann hierarchy. Equations for a few lowest orders are 
given in Appendix A. In order to numerically solve this system of equations, we need to choose the 
highest multipole order $l_{max}$ at which to close the hierarchy. Terminating the hierarchy at some 
finite order carries a risk of transferring artificial power back to lower multipoles \citep{ma95, hu95}. 
We tested our results for a number of different boundary conditions for $l_{max}\sim 10$ and found 
that the solutions for the lowest orders ($l=0$ and $l=1$) are almost insensitive to the change in 
the boundary condition, so we choose to set $\delta f_{l_{max}+1}=0$.

\subsection{Unperturbed Background Solution}

\begin{figure*}
\centering
\includegraphics[width=0.47\textwidth]{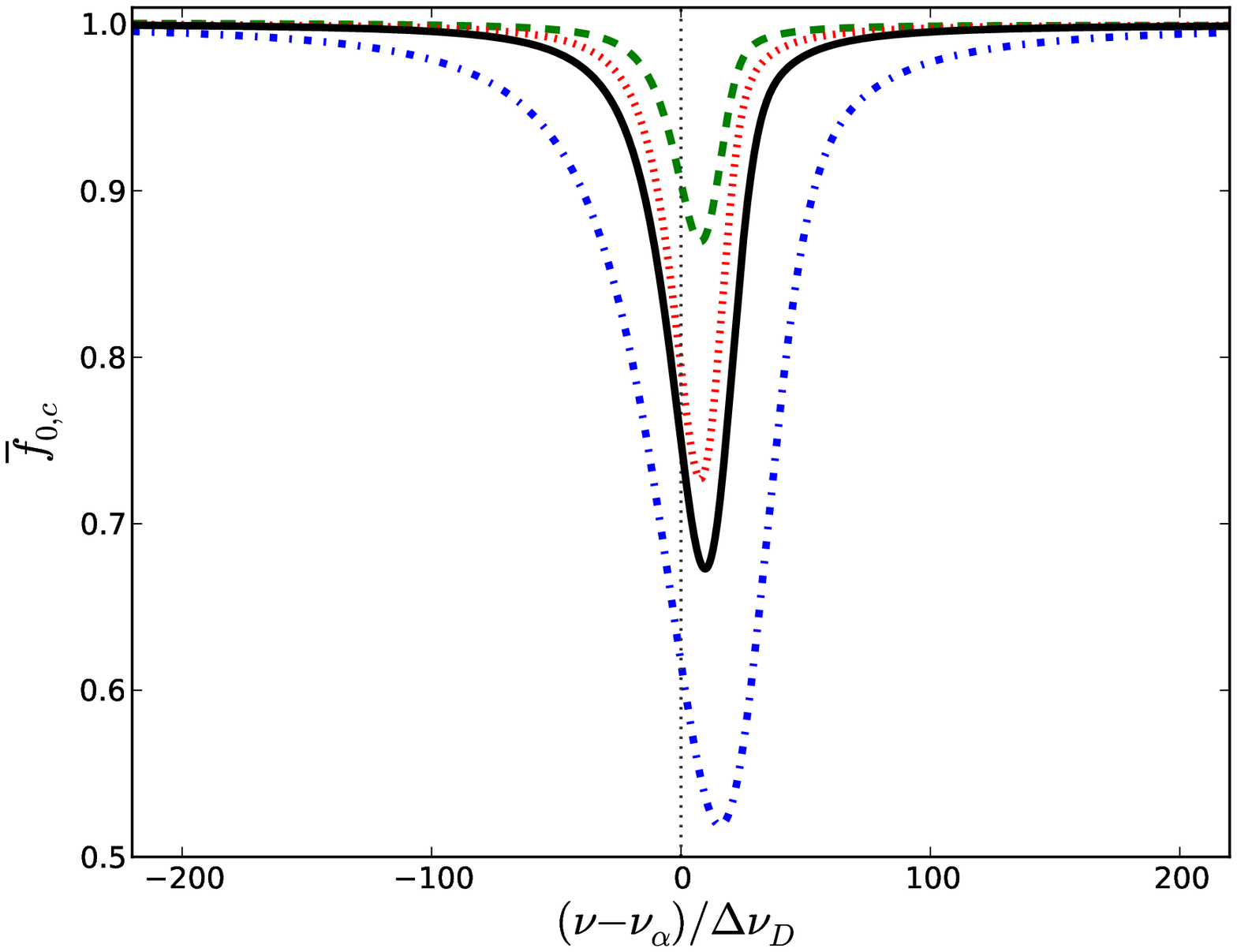}
\includegraphics[width=0.47\textwidth]{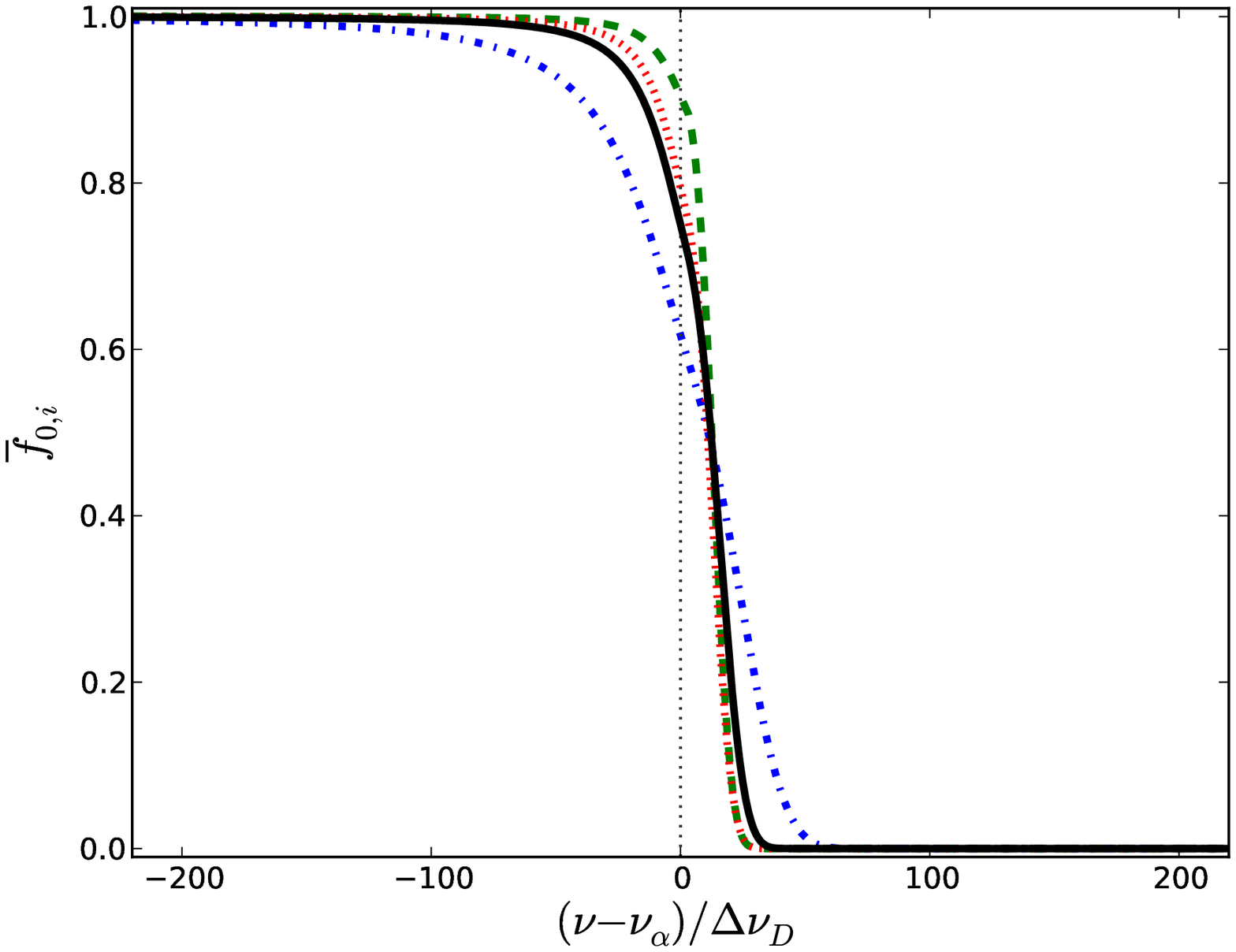}
\caption{Left: background radiation spectrum of the continuum photons around the Ly$\alpha$ frequency 
for the mean (unheated) temperature of $\overline{T}=10$~K and the mean density at $z=20$ (solid black line), 
normalized to the intensity of photons far away from the Ly$\alpha$ frequency. The spectrum shows an asymmetric 
absorption feature that results from combined contributions of scattering diffusivity and atomic recoil. 
Additional lines show how changing the conditions in the gas can modify the absorption feature: the green dashed 
line represents the solution for a five times higher mean temperature, the dash-dotted blue line is obtained for 
a five times higher mean density, and the red dotted line corresponds to the gas of an unaltered mean temperature and 
density, but with a significant velocity divergence. Right: same as on the left, but for the case of 
the injected photons.}
\label{fig:bkg}
\end{figure*}

To obtain the mean background solution $\overline{f}_{0\nu}$, we solve the unperturbed monopole 
equation (i.e., the equation for $l=0$ without any perturbative terms). For the continuum photons, 
the unperturbed equation is
\begin{eqnarray}
\nonumber H\nu_{\alpha}\frac{\partial \overline{f}_{0\nu}}{\partial \nu} +\frac{1}{\nu_{\alpha}^2}\frac{\partial}{\partial \nu} \left[ \nu^2D_{\nu}\left(\frac{\partial \overline{f}_{0\nu}}{\partial \nu}+\frac{h}{k_BT}\overline{f}_{0\nu}\right)\right]=0 \\
\label{eq:unperteq}
\end{eqnarray}
In solving this equation, we follow the procedure described in \cite{cme04}. The resulting spectrum 
(Figure \ref{fig:bkg}, left panel), normalized to the intensity of photons on the blue side far away 
from the line center, shows an asymmetric absorption feature around the Ly$\alpha$ frequency. The shape of the 
feature is determined by the photon drift and diffusion in frequency caused by the scattering off of 
hydrogen atoms. This suppression in the radiation spectrum remains fixed once a steady state has 
been reached; it does not redshift away with the expansion of the universe, indicating energy 
transfer from the radiation field to the gas. 

As shown in the left panel of Figure \ref{fig:bkg}, the absorption feature is deeper for the gas of
higher mean density (shown in blue dash-dotted line) because the scattering rate increases if there 
are more hydrogen atoms present. The feature is shallower for the gas of higher mean temperature 
(green dashed line) because in that case the energy transferred via recoils makes a smaller fraction 
of the average kinetic energy of the atoms. Similarly, the absorption feature is shallower for the 
case of non-zero atomic velocity divergence (red dotted line), which can be thought of as a bulk 
contribution to the kinetic energy of atoms in addition to their thermal motion. This explains 
why the feature changes in the same way as for an increase in temperature.

For the injected photons, the background equation has an extra term $\Psi = H\nu_{\alpha}\delta(\nu_{\alpha})$,
resulting in a different spectral shape (Figure \ref{fig:bkg}, right panel). If there were no
scatterings, the photons would be injected at the Ly$\alpha$ frequency, and they would simply 
redshift to lower frequencies, creating a spectrum shaped as a step function. However, diffusion 
in frequency induced by scattering transfers some of the photons from the red side of the line 
to the blue side. This transfer is enhanced for the gas of higher mean density due to an increased 
scattering rate. Increasing the mean temperature of the gas and introducing bulk motions have 
the opposite effect, as in the case of the continuum photons. Injected photons cause cooling of 
the gas, as the upscattering of photons to the blue side extracts energy from it.

\subsection{Perturbations}

Perturbative terms in our equations have one of the following elements: perturbations to the photon distribution 
function $\delta f_{l\nu}$, non-zero velocity divergence $\delta_{\Theta}$, or perturbations to the 
diffusivity parameter $\delta_D$, which include density and temperature perturbations, $\delta_n$ and 
$\delta_T$, respectively. The full system of coupled differential equations including perturbative terms 
up to linear order is given in Appendix A. We numerically solve it to obtain spectra of perturbations 
for all multipole orders of interest. More details on the numerical implementation can be found in 
Appendix B. Figure \ref{fig:del} shows the lowest two orders ($l=0$ and $l=1$) in the expansion of the 
photon distribution function for different types of perturbations. Spectral features that arise for 
perturbations with wavenumbers in the range considered in this paper have characteristic widths that are 
on the order of several $\Delta\nu_D$ or greater, justifying the use of the Fokker-Planck approximation.

The resulting spectrum for $\delta f_0$ represents photons that are added (or subtracted, depending on 
whether $\delta f_0$ is positive or negative) to the mean solution $\overline{f}_0$ because of a small 
change in the gas temperature, density, or velocity. Since we are considering the photon phase space 
density in the frame of the gas, the dipole term $\delta f_1$ represents the photon flux into or out of 
a Lagrangian region of interest. It vanishes at the line center ($\nu \approx \nu_{\alpha}$) due to a 
very small mean free path of photons near the resonant frequency. 

\begin{figure*}
\centering
\includegraphics[width=0.49\textwidth]{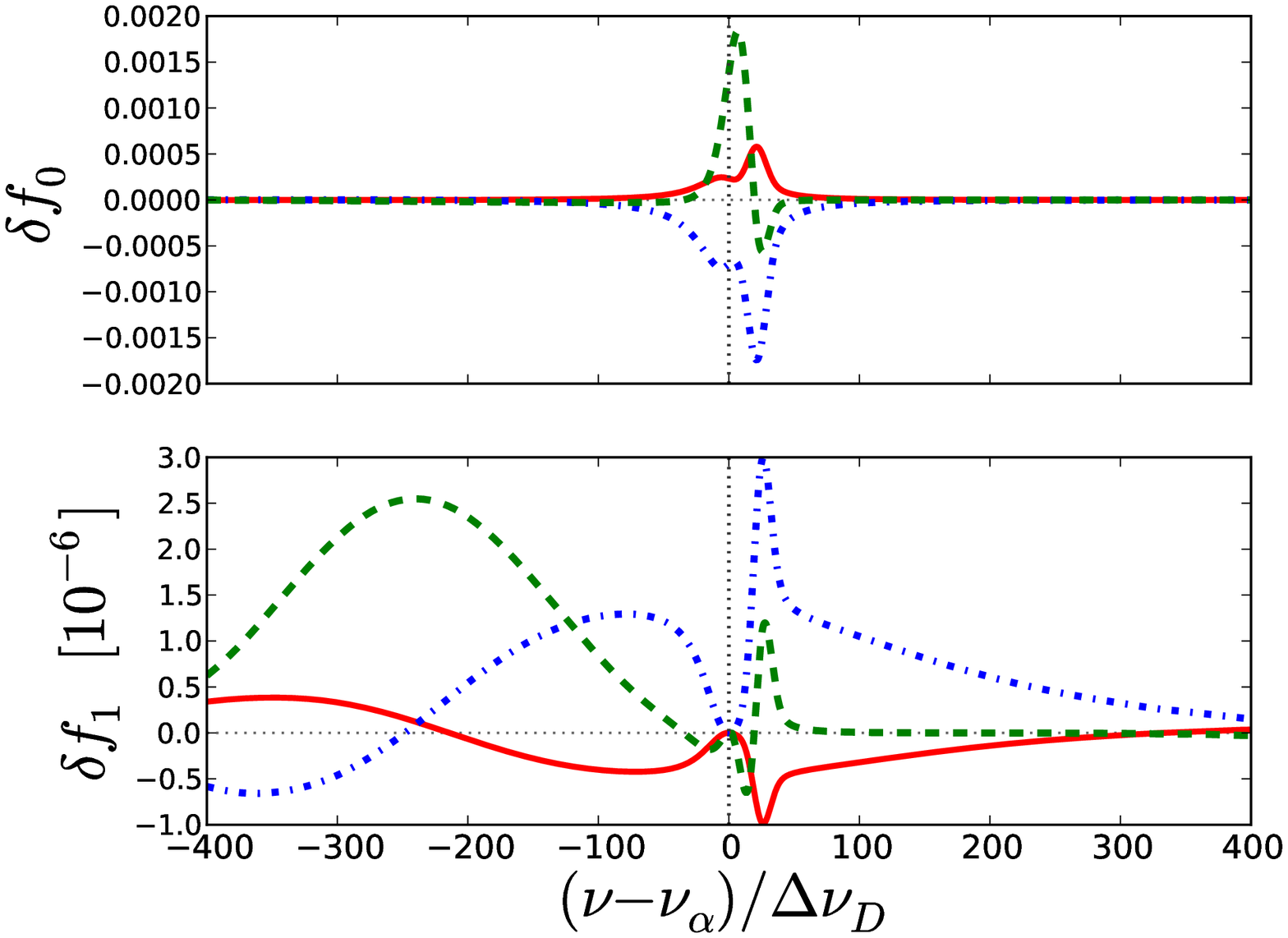}
\includegraphics[width=0.49\textwidth]{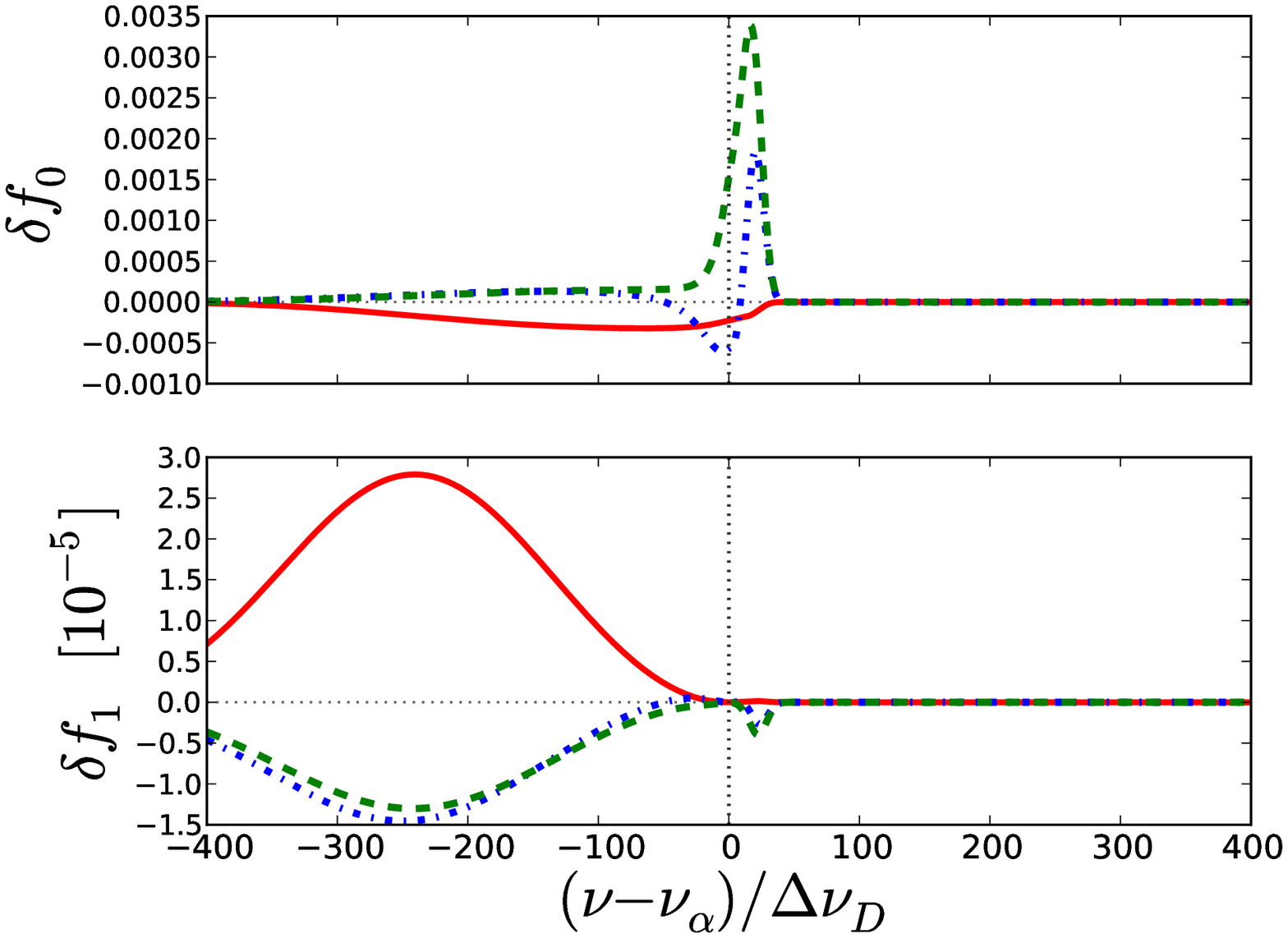}
\caption{Monopole and dipole terms of the perturbed radiation field for the continuum (left) and injected 
(right) photons, shown with the same normalization as in Figure \ref{fig:bkg}. Different curves show the 
results obtained for different types of perturbations, all with the wavenumber $k=1$~cMpc$^{-1}$: green 
dashed curves correspond to $1\%$ temperature perturbations ($\delta_T=0.01$), blue dash-dotted lines 
represent $1\%$ perturbations in the density ($\delta_n =0.01$), and red solid lines show the results 
for introducing $1\%$ perturbation in the velocity divergence ($\delta_{\Theta}/H=0.01$) for the continuum 
photons and 10 times smaller perturbation in size and amplitude for the case of the injected photons.}
\label{fig:del}
\end{figure*}

\section{Heating Rates}
\label{sec:hrate}

In the case of the continuum photons, the gas and the radiation field form a closed system whose energy is
conserved. The rate at which the gas is heated thus equals the negative time change of the radiation energy: 
\begin{equation}
\left[\frac{\partial U_{gas}}{\partial t} + \frac{\partial U_{rad}}{\partial t} \right]_{c}= 0 \ \mbox{,}
\end{equation}
where $U$ is used to denote the energy density (in erg~cm$^{-3}$) of the gas and radiation. The subscript $c$ indicates
that this equation holds for the continuum photons.
For the injected photons, however, there is an external source of energy that needs to be taken into account:
\begin{equation}
\left[\frac{\partial U_{gas}}{\partial t} + \frac{\partial U_{rad}}{\partial t} \right]_{i}= h\nu_{\alpha}\dot{N}_{i} \ \mbox{,}
\end{equation}
where $\dot{N}_{i}$ is the generation rate of the injected photons.

The radiation energy density is given by 
\begin{equation}
U_{rad} = \int n_{\nu} h\nu d\nu\ \ \mbox{.}
\label{eq:energdens}
\end{equation}
The number density of photons of frequency $\nu$ is $n_{\nu}$ (in units of cm$^{-3}$~Hz$^{-1}$).
It is related to the specific intensity $J_{\nu}$ (intensity by the number of photons, not their energy,
given in units of cm$^{-2}$~s$^{-1}$~Hz$^{-1}$~sr$^{-1}$) through the following relation
\begin{equation}
n_{\nu}= \frac{4\pi J_{\nu}}{c} = \frac{8\pi \nu^2 f_{0\nu}}{c^3} \ \mbox{.}
\end{equation}
Thus the photon energy density takes the form
\begin{equation}
U_{rad} = \frac{8\pi h}{c^3}\int \nu^3f_{0\nu}d\nu \ \ \mbox{,}
\label{eq:energdens2}
\end{equation}
where the integral needs to be taken over a wide enough range around the Ly$\alpha$ frequency to include all 
significant spectral features.
The gas heating rate per unit volume, given in units of erg~cm$^{-3}$~s$^{-1}$, is the rate of change of the 
gas energy density $\Gamma \equiv \partial U_{gas}/\partial t $. For the sake of brevity, from now on we refer 
to $\Gamma$ simply as the heating rate. 
For the continuum photons, the heating rate is given by 
\begin{equation}
\Gamma_c= -\frac{\partial U_{rad}}{\partial t}\bigg|_c = -\frac{8\pi h}{c^3}\int \nu^3\frac{\partial f_{0\nu}}{\partial t}\bigg|_cd\nu \ \mbox{.}
\end{equation}
For the injected photons, there is an additional term
\begin{equation}
\Gamma_i=  -\frac{8\pi h}{c^3}\int \nu^3\frac{\partial f_{0\nu}}{\partial t}\bigg|_id\nu +  h\nu_{\alpha}\dot{N}_i \mbox{.}
\label{eq:gammai}
\end{equation}
In the first term on the right side, we can make use of the expression for the time derivative of $f_{0\nu}$ given in Equation (\ref{eq:fpap}).
Note that in the case of the injected photons, the source function $\Psi$ is non-zero.
The photon injection rate in Equation (\ref{eq:gammai}) can be written as
\begin{eqnarray}
\nonumber \frac{\partial {N}_i}{\partial t} &=& \frac{\partial}{\partial t}\left( \int n_{i,\nu} d\nu \right) = \frac{\partial}{\partial t}\left( \int \frac{8\pi\nu^2}{c^3} f_i d\nu \right) \\
&=&\int \frac{8\pi\nu^2}{c^3} \frac{\partial f_i }{\partial t}d\nu  = \int \frac{8\pi\nu^2}{c^3} \Psi d\nu
\end{eqnarray}
The two terms containing the source function cancel out - the injection of photons does not contribute to the gas heating rate. 
The heating of the gas is caused solely by the frequency diffusivity part of the collision term given by Equation (\ref{eq:fpap}).

The largest contribution to the gas heating rate comes from the part of the spectrum around the 
line center, thus it is convenient to separate the radiation energy density in the following way:
\begin{equation}
U = \frac{8\pi h}{c^3}\left[\nu_{\alpha}\int_{\nu_1}^{\nu_2} \nu^2 f_{0\nu}d\nu + \int_{\nu_1}^{\nu_2} (\nu-\nu_{\alpha})\nu^2 f_{0\nu}d\nu\right] \mbox{.}
\end{equation}
The corresponding heating rate is
\begin{equation}
\Gamma =-\frac{8\pi h}{c^3}\biggl[\nu_{\alpha}\int_{\nu_1}^{\nu_2} \nu^2 \frac{df_{0\nu}^{coll}}{dt}d\nu + \int_{\nu_1}^{\nu_2} (\nu-\nu_{\alpha})\nu^2 \frac{df_{0\nu}^{coll}}{dt}d\nu\biggr] \mbox{.}
\end{equation}
Using the expression given in Equation (\ref{eq:fpap}), the first term on the right side becomes
\begin{eqnarray}
 \nonumber &-&\frac{8\pi h\nu_{\alpha}}{c^3}\int_{\nu_1}^{\nu_2} \frac{\partial}{\partial\nu}\left[ \nu^2 D_{\nu}\left( \frac{\partial f_{0\nu}}{\partial\nu} + \frac{h}{k_B T}f_{0\nu}\right) \right] d\nu \\
&=& -\frac{8\pi h\nu_{\alpha}}{c^3}\left[ \nu^2 D_{\nu}\left( \frac{\partial f_{0\nu}}{\partial\nu} + \frac{h}{k_B T}f_{0\nu}\right) \right]\bigg|_{\nu_1}^{\nu_2} \ \mbox{.}
\end{eqnarray}
The contribution of this term to the total heating rate is vanishingly small because $D_{\nu}$ approaches 
zero far from the line center. The remaining term is
\begin{equation}
\nonumber \Gamma = -\frac{8\pi h}{c^3}\int_{\nu_1}^{\nu_2} (\nu-\nu_{\alpha})\frac{\partial}{\partial\nu}\left[ \nu^2 D_{\nu}\left( \frac{\partial f_{0\nu}}{\partial\nu} + \frac{h}{k_B T}f_{0\nu}\right) \right]d\nu
\end{equation}
We can separate the heating rate into the contribution of the mean background radiation field and
the contribution of the perturbations with
\begin{equation}
\Gamma = \overline{\Gamma} + e^{ikx_3}\delta \Gamma \ \mbox{,}
\end{equation}
where $\overline{\Gamma}$ and $\delta \Gamma$ represent the background and perturbation heating, respectively. 
Making use of Equation (\ref{eq:unperteq}) for the background heating rate, and an analogous expression for the case 
of perturbations, obtained from Equation (\ref{eq:fsfull}), we get
\begin{eqnarray}
\nonumber \overline{\Gamma} &=& \frac{8\pi h}{c^3}\int_{\nu_1}^{\nu_2} (\nu-\nu_{\alpha})H\nu_{\alpha}^3 \frac{\partial \overline{f}_{0\nu}}{\partial\nu} d\nu \\
\nonumber &=& -\frac{8\pi h \nu_{\alpha}^3}{c^3}H(\nu-\nu_{\alpha})(\overline{f}_{\alpha}-\overline{f}_{0\nu})\bigg|_{\nu_1}^{\nu_2}\\
& & + \frac{8\pi h \nu_{\alpha}^3}{c^3}H\int_{\nu_1}^{\nu_2} (\overline{f}_{\alpha}-\overline{f}_{0\nu})d\nu \ \mbox{,}
\label{eq:hrcentermean}\\
\nonumber \delta \Gamma &=& \frac{8\pi h}{c^3}\int_{\nu_1}^{\nu_2} \left(H\nu_{\alpha}^3 \frac{\partial \delta f_{0\nu}}{\partial\nu} + \frac{\nu_{\alpha}^3\delta_{\Theta}}{3}\frac{\partial \overline{f}_{0\nu}}{\partial \nu} + \nu_{\alpha}^2 kc \delta f_{1\nu}\right)\\
\nonumber & & \times (\nu-\nu_{\alpha})d\nu \\
\nonumber &=& \frac{8\pi h \nu_{\alpha}^3}{c^3}H\left[(\nu-\nu_{\alpha})\delta f_{\alpha}\bigg|_{\nu_1}^{\nu_2} -\int_{\nu_1}^{\nu_2} \delta f_{0\nu}d\nu \right]\\
\nonumber & & - \frac{8\pi h \nu_{\alpha}^3\delta_{\Theta}}{3c^3}\left(\nu-\nu_{\alpha}\right)\left( \overline{f}_{\alpha}-\overline{f}_{0\nu} \right)\bigg|_{\nu_1}^{\nu_2} \\
\nonumber & & + \frac{8\pi h \nu_{\alpha}^3\delta_{\Theta}}{3c^3}\int_{\nu_1}^{\nu_2} \left( \overline{f}_{\alpha}-\overline{f}_{0\nu} \right) d\nu \\
& & +  \frac{8\pi h \nu_{\alpha}^2}{c^3}kc\int_{\nu_1}^{\nu_2} (\nu-\nu_{\alpha})\delta f_{1\nu}d\nu \ \mbox{,}
\label{eq:hrcenterpert}
\end{eqnarray}

The above formulae are appropriate for frequencies around the line center. However, they might cause significant
numerical errors in the wings of the line due to approximations made in deriving them. Hence, we
use another expression to evaluate the heating rate in the wings:
\begin{eqnarray}
\nonumber \Gamma &=&  -\frac{8\pi h}{c^3}\left[ (\nu-\nu_{\alpha})\nu^2D_{\nu}\left( \frac{\partial f_{0\nu}}{\partial\nu} + \frac{h}{k_B T}f_{0\nu}\right)\bigg|_{\nu_1}^{\nu_2}\right] \\
&& + \frac{8\pi h}{c^3}\left[\int_{\nu_1}^{\nu_2}\nu^2D_{\nu}\left( \frac{\partial f_{0\nu}}{\partial\nu} + \frac{h}{k_B T}f_{0\nu}\right)d\nu\right] \ \mbox{.}
\label{eq:hrwings}
\end{eqnarray}

We report the calculated heating rates in terms of a dimensionless quantity, which we call the \textit{relative heating}, 
that measures the energy transferred to the gas per Hubble time, relative to the thermal energy of the gas ($3\overline{n}k_B \overline{T}/2$):
\begin{equation}
\frac{\Gamma }{\frac{3}{2}\overline{n}k_B \overline{T}H(z)} = \frac{\overline{\Gamma}}{\frac{3}{2}\overline{n}k_B \overline{T}H(z)} + \frac{\delta \Gamma e^{ikx_3}}{\frac{3}{2}\overline{n}k_B \overline{T}H(z)}  \ \ \mbox{,}
\label{eq:thermalnerg}
\end{equation}
where $\overline{n}$ is the number density of all baryons, not just hydrogen atoms. Contributions of 
density, temperature, and velocity perturbations to the heating rate are incorporated into $\delta \Gamma$ 
and can be treated independently for each type of perturbation:
\begin{eqnarray}
\delta \Gamma &=& \frac{\partial \Gamma}{\partial n}\delta_n + \frac{\partial \Gamma}{\partial T}\delta_T + \frac{\partial \Gamma}{\partial \Theta}\delta_{\Theta} \ \ \mbox{.}
\end{eqnarray}
Hence, the perturbative part of Equation (\ref{eq:thermalnerg}) is given by
\begin{equation}
\frac{\delta \Gamma e^{ikx_3}}{\frac{3}{2}\overline{n}k_B \overline{T}H(z)}= \left[ C_n \delta_n + C_T \delta_T + C_{\Theta} \frac{\delta_{\Theta}}{H}\right]\frac{\overline{J}_{\alpha}}{\tilde{J}_0}e^{ikx_3} \ \ \mbox{,}
\label{eq:diffthenerg}
\end{equation}
where  $\overline{J}_{\alpha}$ is the specific intensity of incoming photons and 
\begin{equation}
\tilde{J}_0 = \frac{n_H c}{4\pi \nu_{\alpha}} = \frac{2\nu_{\alpha}^2 \tilde{f}_0 }{c^2} 
\end{equation}
is the intensity corresponding to one photon per frequency octave per hydrogen atom in the 
universe \citep{cme04}. In terms of energy intensity, this corresponds to 
$\tilde{J}_0 h\nu_{\alpha} \approx 2.5\times 10^{-20}$~erg~cm$^{-2}$~s$^{-1}$~sr$^{-1}$~Hz$^{-1}$ 
at $z= 20$. In the model of \cite{ciardi03}, the intensity of Ly$\alpha$ background at 
$z\sim 20$ is on the order of $10^{-20}$~erg~cm$^{-2}$~s$^{-1}$~sr$^{-1}$~Hz$^{-1}$, making 
$\overline{J}_{\alpha} / \tilde{J}_0$ a factor of order unity. In general, one expects it to be a rapidly
increasing function of redshift. Since it takes $\ge 1$ H-ionizing ($\nu>\frac43\nu_\alpha$) photons
per atom to ionize the universe, and since the non-ionizing photons that redshift into Lyman series lines
do not suffer from absorption in the emitting galaxies, we expect that $\overline J_\alpha/\tilde J_0$
should reach unity at an early stage of reionization \citep{cme04}.

We have defined dimensionless heating coefficients $C$ for all three types of perturbations by
\begin{eqnarray}
 C_n &=& \frac{2\tilde{f}_0}{3\overline{n}k_B\overline{T}H}\frac{\partial \Gamma}{\partial n} \ \mbox{,} 
\label{hcoef1} \\
 C_T &=& \frac{2\tilde{f}_0}{3\overline{n}k_B\overline{T}H}\frac{\partial \Gamma}{\partial T}  \ \mbox{,~and}
\label{hcoef2} \\
 C_{\Theta} &=& \frac{2\tilde{f}_0}{3\overline{n}k_B\overline{T}}\frac{\partial \Gamma}{\partial \Theta} \ \mbox{;}
\label{hcoef3}
\end{eqnarray}
these represent the heat input per Hubble time in units of the thermal energy of the gas, if $\overline J_\alpha/\tilde J_0=1$.

\section{Results and Discussion}
\label{sec:results}

We perform calculations described in the previous sections for neutral hydrogen gas ($x_{HI}=1$) 
at the redshift of $z=20$. The mean baryon number density is easily obtained from the current 
baryon density of the universe $\overline{n}_0 = 2.5\times 10^{-7}$~cm$^{-3}$ (WMAP-9 result, 
\citealt{ben12}) as $\overline{n}(z) = \overline{n}_0(1+z)^3$. Finally, to get the number density 
of hydrogen atoms we take into account that over $90\%$ (by number) of the baryonic content is 
in the form of hydrogen atoms. For the mean temperature of the gas we take the value 
$\overline{T}=10$~K (see Figure 1 in \citealt{prfu06}).

\subsection{Heating from Unperturbed Radiation}

The calculated contribution to the relative heating coming from the mean (unperturbed) background photons can be expressed as
\begin{equation}
\frac{\overline{\Gamma}_c}{\frac{3}{2}\overline{n}k_B\overline{T}H} = 0.13\frac{\overline{J}_{\alpha , c}}{{J}_0} \
\end{equation}
for the continuum photons, and as
\begin{equation}
\frac{\overline{\Gamma}_i}{\frac{3}{2}\overline{n}k_B\overline{T}H} = -0.07\frac{\overline{J}_{\alpha , i}}{{J}_0}
\end{equation}
for the injected photons. The relative heating caused by the scattering of the injected photons is 
negative, indicating that the gas is cooled down by this interaction. At this temperature and density, 
heating of the gas caused by the scattering of the continuum photons prevails over cooling by the 
injected photons, not only because the effect itself is slightly stronger, but also because the flux 
of the injected photons is smaller than the flux of the continuum photons; the ratio of the injected 
and continuum photons is around 10$\%$-20$\%$ \citep{prfu06, chsh07}.

\subsection{Heating from Perturbations}

\begin{figure}
\centering
\includegraphics[width=0.49\textwidth]{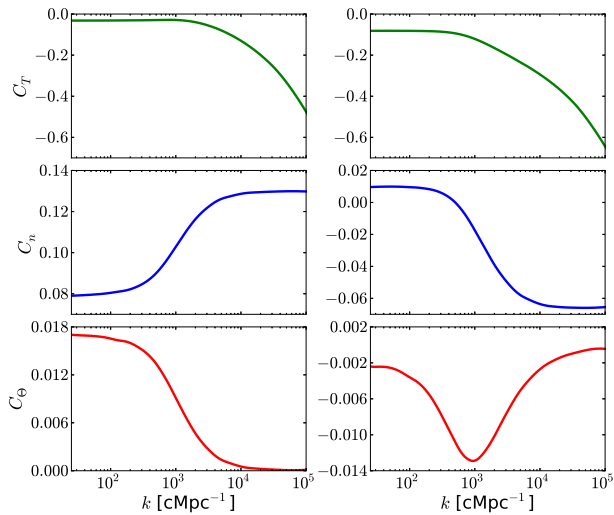}
\caption{Heating coefficients, defined by expressions (\ref{hcoef1}) - (\ref{hcoef3}), 
for the continuum (left) and injected photons (right) show contributions of temperature, density, and 
velocity perturbations (from top to bottom) to the total heating rate as functions of the perturbation 
wavenumber $k$, given in comoving units.}
\label{fig:Ccoef}
\end{figure}

A new result of this study is the additional contribution to the relative heating caused by inhomogeneities
in the gas. The differential relative heating due to perturbations is given by Equation (\ref{eq:diffthenerg}).
The values of all three heating coefficients $C$ for perturbations of different wavenumbers $k$ are shown 
in Figure \ref{fig:Ccoef}.

\begin{figure}
\centering
\includegraphics[width=0.48\textwidth]{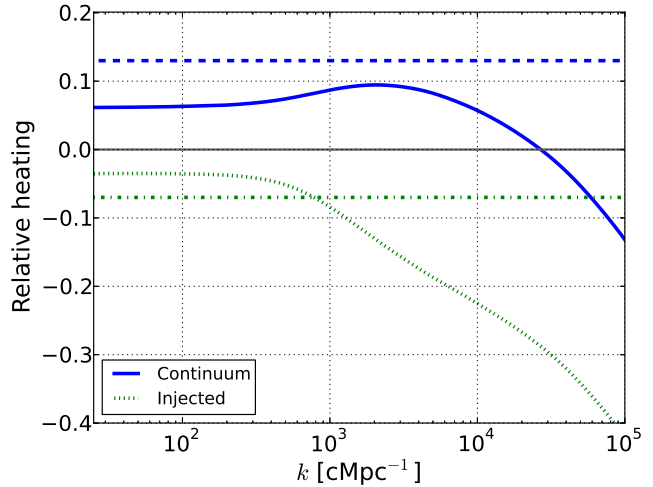}
\caption{Solid and dotted lines show the joint contribution of the temperature and density perturbations 
to the relative heating for the continuum and injected photons, respectively, as a function of perturbation 
wavenumber (comoving). These values need to be multiplied by the amplitude of the density perturbations to 
give the perturbative heating that can then be compared to the unperturbed effect, shown in dashed and dash-dotted lines. 
On very small scales, the perturbative heating from the continuum photons changes sign relative to 
the mean effect, causing damping of small scale perturbations. The same does not occur for the injected 
photons - they cause negative heating (i.e., cooling) of gas on all scales. Their contribution relative 
to the continuum photons is diminished by the fact that their intensity is lower, making only $\sim 10\%-20\%$ 
of the intensity of the continuum photons.}
\label{fig:adpert}
\end{figure}

On large scales, corresponding to small values of the wavenumber $k$, having a positive perturbation in
the temperature or density is similar to having a region with no perturbations, but with an increased
mean value. As shown in Figure \ref{fig:bkg} and discussed in Section \ref{sec:radtrans}, an increase
in $\overline{T}$ causes the absorption feature in the spectrum of $\overline{f}_{0}$ to be shallower,
whereas an increase in $\overline{n}$ makes the feature deeper. The heating is proportional to the
integral of the difference between the spectrum without any scattering (which is just a flat spectrum
for the case of the continuum photons) and the real spectrum, hence it is proportional to the area of
the absorption feature. Therefore, the heating will be smaller in the case of higher $\overline{T}$,
which is why $C_T$ has negative values on large scales. On the other hand, $C_n$ is positive, since an
increase in $\overline{n}$ causes a larger heating. The absolute values of heating coefficients 
increase for larger values of $k$, indicating that these effects are enhanced on smaller scales in
the case of the continuum photons.

To estimate the contribution of perturbations to the total heating of the gas, we need to know the relative
strengths of different kinds of perturbations, in addition to the values of the heating coefficients.
\cite{naoz05} calculated the ratio of temperature and density perturbations as a function of wavenumber 
$k$ at several redshifts, including $z=20$, the epoch that we consider in this work. The ratio of 
$\delta_T$ and $\delta_n$ at $z=20$ is almost constant for small-scale perturbations, 
$\delta_T/\delta_n \approx 0.55$. We make use of this result to show (Figure \ref{fig:adpert}) the
joint contribution of the density and temperature perturbations ($C_n + 0.55C_T$) relative to the 
heating caused by the unperturbed background photons. We ignore perturbations in the velocity 
because the magnitude of $C_{\Theta}$ is much smaller than that of $C_T$ or $C_n$ for most wavenumbers.

On large scales (i.e., small values of $k$), the heating caused by perturbations has the same sign as 
the mean effect: positive for the continuum photons and negative for the injected photons, which indicates a 
small increase in the gas heating by the continuum photons in regions of higher density and temperature 
due to the increased scattering rate. Cooling by the injected photons in those regions will also increase. 
These additional contributions to the heating are very small. The values shown in Figure \ref{fig:adpert}, 
which are already a factor of $\sim 2$ smaller than the mean effect, need to be multiplied by the amplitude 
of density perturbations to give the actual relative heating. Since our formalism is based on the linear 
perturbation theory, it is applicable to perturbations of the order of a few percent or smaller. Hence, the 
additional heating at large scales can only make a few percent of the mean (unperturbed) Ly$\alpha$ heating.

An interesting feature appears for perturbations on very small scales ($k \sim 2\times 10^4$~cMpc$^{-1}$).
For the case of the continuum photons, values of the relative heating, shown in Figure \ref{fig:adpert}, 
turn from positive to negative. On length scales below that threshold, perturbations act in the opposite 
direction from the unperturbed effect; they reduce the heating of the gas in positively perturbed regions 
(i.e., regions of higher density and temperature than the mean). The opposite happens for cooler and 
underdense regions. The effect of the injected photons remains unchanged. Hence, perturbations on scales 
smaller than that corresponding to $k \sim 2\times 10^4$~cMpc$^{-1}$ will be damped due to the effect of 
thermal conduction by Ly$\alpha$ photons. This length scale, however, is roughly two orders of magnitude 
smaller than the Jeans scale.

For gas of higher mean temperature, e.g., with $\overline{T}=20$~K, the result stays qualitatively the same, 
only the values of the relative heating, both the mean effect and the contribution of perturbations, are reduced.

\section{Conclusions}
\label{sec:conclusions}

The resonant scattering of Ly$\alpha$ photons produced by early generations of luminous objects can cause moderate
heating of high-redshift IGM. Ly$\alpha$ photons and hydrogen atoms exchange energy during scattering due to
atomic recoil. Details of radiative transfer are further complicated by the frequency diffusion of photons caused by 
scattering and the drift to lower frequencies due to Hubble expansion. In the optically thick limit, the gas 
and the radiation field approach statistical equilibrium, which greatly reduces the energy exchange. Taking into 
account all of these effects, an asymmetric absorption feature is created in the radiation spectrum around the 
Ly$\alpha$ frequency in the case of the continuum photons that redshift directly into Ly$\alpha$ from the blue 
side of the line. Photons that redshift into higher resonances and then cascade into Ly$\alpha$ are called the 
injected photons. Their spectrum has a shape of a modified step function around the Ly$\alpha$ frequency. 
Scattering of the continuum photons causes heating of the IGM proportional to the area of the absorption 
feature. This heating is higher for gas of lower mean temperature and higher density. The injected photons, on the other 
hand, cause cooling of the gas because they preferentially scatter off atoms moving in the opposite direction. 

In this paper, we study the effect of Ly$\alpha$ scattering on high-redshift ($z=20$) IGM with linear perturbations 
in density, temperature, and velocity divergence. We are primarily interested in small-scale perturbations that can
be affected by thermal conduction via Ly$\alpha$ photons. For perturbations with scales smaller than the Ly$\alpha$ 
diffusion length-scale, photons can diffuse into regions where they are further away from being in a statistical 
equilibrium with the gas, causing enhancement in the energy exchange. To find the exact scale at which this occurs,
we solve radiative transfer equations numerically, using the Fokker-Planck approximation.

We find that the scale at which this effect becomes relevant is very small, corresponding to a comoving wavenumber 
of $k \sim 2\times 10^4$~Mpc$^{-1}$, which is a factor of $\sim 100$ smaller than the Jeans scale. On larger scales,
where structures in the IGM are expected to be present, the heating perturbations add a correction to the mean 
effect that is on the order of the amplitude of the density or temperature perturbation in the gas. Since our formalism 
is based on the linear perturbation theory, this makes only a few percent difference to the gas heating in typical cases.

\acknowledgments We thank Tzu-Ching Chang and Matthew Schenker for helpful comments and Tejaswi Venumadhav for useful 
conversations. A.O. acknowledges support from the International Fulbright Science \& Technology Award. During the 
preparation of this paper, C.H. has been supported by the U.S. Department of Energy under contract DE-FG03- 02-ER40701, 
the Alfred P. Sloan Foundation, and the David and Lucile Packard Foundation.

{}

\appendix
\section{System of equations}

Equations for different multipoles of the perturbed radiation field are given by:
\begin{eqnarray}
\nonumber H\nu \frac{\partial \delta f_{0\nu}}{\partial\nu} + \frac{1}{3}\nu\delta_{\Theta} \frac{\partial \overline{f}_{0\nu}}{\partial \nu} &+& kc \delta f_{1\nu} + \frac{\partial \overline{D}_{\nu}}{\partial\nu}\left( \frac{\partial \delta f_{0\nu}}{\partial\nu} + \frac{h}{k_BT}\delta f_{0\nu} \right) +\frac{\partial \overline{D}_{\nu}}{\partial \nu}\delta_D\left( \frac{\partial \overline{f}_{0\nu}}{\partial \nu} + \frac{h}{k_BT} \overline{f}_{0\nu}\right)+\\
&+& \overline{D}_{\nu}\left( \frac{\partial^2 \delta f_{0\nu}}{\partial \nu^2} + \frac{h}{k_BT}\frac{\partial \delta f_{0\nu}}{\partial\nu} \right)+ \overline{D}_{\nu}\delta_D\left( \frac{\partial^2 \overline{f}_{0\nu}}{\partial \nu^2} + \frac{h}{k_BT}\frac{\partial \overline{f}_{0\nu}}{\partial\nu} \right)=0 \ \mbox{,} \ \ \ \mbox{$(l=0)$}
\end{eqnarray}
\begin{eqnarray}
H\nu \frac{\partial \delta f_{1\nu}}{\partial \nu} - \frac{kc}{3}\left(\delta f_{0\nu}-2\delta f_{2\nu}\right) -n_Hc\sigma(\nu,T)\delta f_{1\nu} =0 \ \mbox{,} \ \ \ \mbox{$(l=1)$}
\end{eqnarray}
\begin{eqnarray}
 H\nu\frac{\partial \delta f_{2\nu}}{\partial \nu} - \frac{2}{15}\nu\delta_{\Theta} \frac{\partial \overline{f_{0\nu}}}{\partial\nu} - \frac{kc}{5}\left(2\delta f_{1\nu}-3\delta f_{3\nu}\right) -n_Hc\sigma(\nu,T)\delta f_{2\nu} =0 \ \mbox{,} \ \ \ \mbox{$(l=2)$}
\end{eqnarray}
\begin{eqnarray}
\nonumber &\cdots &\\
H\nu \frac{\partial \delta f_{l_{max}\nu}}{\partial \nu} &-& \frac{kc}{2l_{max}+1}\left(l_{max}\delta f_{\left(l_{max}-1\right)\nu}-(l_{max}+1)\delta f_{\left(l_{max}+1\right)\nu}\right) -n_Hc\sigma(\nu,T)\delta f_{l_{max}\nu}=0 \ \mbox{,} \ \ \ \mbox{$(l=l_{max})$}
\end{eqnarray}
where $k$ is the physical wavenumber, not comoving. In the first equation, $\overline{D}_{\nu}$ 
is used to denote the mean value of the diffusivity parameter. The full expression for this 
quantity, including perturbations, can be written as 
$D_{\nu} = \overline{D}_{\nu}(1+\delta_D) = \overline{D}_{\nu}(1+ \delta_n + \delta_T)$.

\section{Numerical calculations}

To solve the above described system of equations, we truncate the series at $l_{max} = 8$ by setting 
$\delta f_{9\nu}=0$. The choice of the largest considered multipole could of course be different. 
In our analysis, we only use the solutions for the monopole and dipole terms, but we want to keep 
as many higher orders as possible to avoid introducing significant errors into the lowest multipoles 
by making an artificial truncation of the series too close to them. On the other hand, tracking too 
many multipole orders becomes computationally challenging. We find that keeping nine multipoles is 
optimal: the computation can be done in a reasonable amount of time and increasing $l_{max}$ 
by one changes the resulting heating rate by $2\%$ at most (in most cases much less than that).

Once we have a finite set of differential equations, we create an equidistant frequency 
grid containing a large number ($7\times 10^5$) of frequencies centered at $\nu_{\alpha}$. The frequency 
range covered in our grid spans $3.3\times 10^{12}$~Hz (corresponding to $\sim 1000\ \Delta\nu_D$) on 
each side of the Ly$\alpha$ frequency. These numbers could have been chosen differently without 
significantly affecting the final result. For example, decreasing the frequency range by $10\%$ changes 
the computed heating rate by no more than $1\%-2\%$. We can change the size of the frequency grid to 
test the convergence of our method. Increasing the number of frequencies in the grid spanning the 
same frequency range to $8\times 10^5$ changes the result by $\sim 2\%$ or less. 

In order to avoid numerical errors that can occur near the boundaries of the grid, in our calculations 
we do not take into account outer $3\times 10^4$ frequencies at both ends of the grid. As mentioned in 
Section \ref{sec:hrate}, to calculate the heating rate we use Equations (\ref{eq:hrcentermean}) and 
(\ref{eq:hrcenterpert}) in the central 
part of the grid and Equation (\ref{eq:hrwings}) in the outer parts. The exact number of frequencies included 
in these outer parts, the so-called wings, does not significantly affect the result, as long as the central 
spectral features, contained within $\sim 100 \Delta\nu_D$ of the Ly$\alpha$ line center, are not included. 
The difference between having $\sim 20\%$ of frequencies in the wings and having $40\%$ is less than $4\%$.  

For the described grid of $7\times 10^5$ frequencies and with 9 multipoles, our system of equations forms
a matrix of dimension $(9\times 7\times 10^5) \times (9\times 7\times 10^5)$. Manipulating such a large
matrix can be challenging. Fortunately, most of the elements of this matrix are zero, hence we make use 
of SciPy sparse matrix package (scipy.sparse) to construct the matrix and solve the sparse linear system.

To approximate differentiation in the equations, we first used the central difference method
\begin{equation}
f^{\prime}(\nu) = \frac{f(\nu +\Delta\nu) - f(\nu -\Delta\nu)}{2\Delta\nu} + \mathcal{O}(\Delta\nu^2)\mbox{,}
\end{equation}
where $\Delta\nu$ is the grid step.
However, due to numerical oscillations that occurred near the end of the frequency grid, we introduced
dissipation in the form of the forward difference contribution
\begin{equation}
f^{\prime}(\nu) = \frac{f(\nu +\Delta\nu) - f(\nu)}{\Delta\nu} + \mathcal{O}(\Delta\nu)\mbox{.}
\end{equation}
Hence, the derivatives are given by a linear combination of the central and forward difference terms
\begin{equation}
f^{\prime}(\nu) = \epsilon\frac{f(\nu +\Delta\nu) - f(\nu)}{\Delta\nu} + \left(1-\epsilon\right)\frac{f(\nu +\Delta\nu) - f(\nu -\Delta\nu)}{2\Delta\nu} \ \mbox{.}
\end{equation}
Parameter $\epsilon$ describes the contribution of the forward difference method, and it changes linearly
from zero at the center to unity at the ends of the frequency grid.

\end{document}